\documentclass{emulateapj}
\usepackage{verbatim} 

\shorttitle{CO in Radio Galaxies}
\shortauthors{Smol\v{c}i\'{c} \& Riechers}

\def\f#1   {Fig.~\ref{#1}}
\def\s#1   {Sec.~\ref{#1}}
\def\tab#1   {Tab.~\ref{#1}}
\def\t#1   {Tab.~\ref{#1}}

\def\comm#1   {{\tt (COMMENT: #1) }}

\def\kms{~km~s$^{\mathrm{-1}}$}

\def\msol              {$\mathrm{M}_{\odot}$}

\def\wh                {W~Hz$^{-1}$}

\def\kms{~km~s$^{\mathrm{-1}}$}

\def\smo               {Smol\v{c}i\'{c}}

\slugcomment{  }

\begin{document}

\title{The Molecular Gas Content of $z<0.1$ Radio Galaxies:\\
  Linking the AGN Accretion Mode to Host Galaxy Properties}

\author{
        V.~Smol\v{c}i\'{c}\altaffilmark{1,2,3} and
	D.~A.~Riechers\altaffilmark{4,5}
        }
\altaffiltext{1}{European Southern Observatory, Karl-Schwarzschild-Strasse 2, 
D-85748 Garching b. M\"unchen, Germany}
\altaffiltext{2}{Argelander Institut for Astronomy, Auf dem H\"{u}gel 71, Bonn, D-53121, Germany}
\altaffiltext{3}{ESO ALMA COFUND Fellow}
\altaffiltext{4}{California Institute of Technology, MC\,249-17, 1200 East
California Boulevard, Pasadena, CA 91125, USA}
\altaffiltext{5}{Hubble Fellow}

\begin{abstract}
  One of the main achievements in modern cosmology is the so-called
  `unified model', which successfully describes most classes of active
  galactic nuclei (AGN) within a single physical scheme. However,
  there is a particular class of radio-luminous AGN that presently
  cannot be explained within this framework -- the `low-excitation'
  radio AGN (LERAGN). Recently, a scenario has been put forward which
  predicts that LERAGN, and their regular `high-excitation' radio AGN
  (HERAGN) counterparts represent different (red sequence vs.\ green
  valley) phases of galaxy evolution. These different evolutionary
  states are also expected to be reflected in their host galaxy
  properties, in particular their cold gas content. To test this, here
  we present CO(1$\rightarrow$0) observations toward a sample of 11 of
  these systems conducted with CARMA. Combining our observations with
  literature data, we derive molecular gas masses (or upper limits)
  for a complete, representative, sample of 21 $z<0.1$ radio AGN.
 Our results yield that HERAGN on average have a factor of $\sim7$
  higher gas masses than LERAGN. We also infer younger stellar ages,
  lower stellar, halo, and central supermassive black masses, as well
  as higher black hole accretion efficiencies in HERAGN relative to
  LERAGN. These findings support the idea that high- and
  low-excitation radio AGN form two physically distinct populations of
  galaxies that reflect different stages of massive galaxy build-up.

\end{abstract}


\section{Introduction}
\label{sec:intro}

Over the past two decades a standard model of AGN has emerged. In this
`unified' model efficient disk accretion of cold matter on the central
supermassive black hole (BH) provides the radiation field that
photoionizes emission-line regions. However, there is a certain
fraction of AGN identified by radio observations that poses a
challenge to the unified model, the so-called low-excitation radio AGN
(hereafter:\ LERAGN).  The main difference between {\em high-excitation
  radio AGN (HERAGN)} and these {\em LERAGN} is that the latter do not
exhibit strong emission lines in their optical spectra (Jackson \&
Rawlings 1997; Evans et al.\ 2006).

Recently, Hardcastle et al.\ (2006) have suggested that high- and
low-excitation radio AGN may represent a principal separator between
populations fundamentally different in their black hole accretion
mechanisms (see also Evans et al. 2006; Allen et al. 2006; Kewley et
al. 2006). They developed a model in which central supermassive black
holes of HERAGN accrete in a standard (radiatively efficient) way from
the cold phase of the intragalactic medium (IGM), while those of
LERAGN are powered in a radiatively inefficient manner by Bondi
accretion of the hot IGM. 
\smo\ (2009) 
showed that low- and high excitation radio AGN exhibit not only
systemic differences in their black hole masses and accretion rate
properties, but also in their host galaxy properties, such as stellar
masses and stellar populations. 
This is consistent with these two classes of radio AGN dividing in a
stellar mass vs.\ color plane in such a way that LERAGN occupy the red
sequence and HERAGN inhabit the so called ``green valley'', a sparsely
populated region between the blue-cloud and the red-sequence (\smo\
2009).

The stellar mass vs.\ color plane can be interpreted as a
time-sequence for galaxy evolution. Galaxies are thought to evolve 
from an initial star-formation-dominated state with blue optical
colors into the most massive “red-and-dead” galaxies through a
transition phase reflected in the green valley (Bell et al. 2004a,
2004b; Borch et al. 2006; Faber et al. 2007; Brown et al. 2007).  In
recent years it has been suggested
that radio outflows from AGN likely play a crucial role in this
massive galaxy build-up \citep{croton06,bower06,sijacki06,sijacki07,fanidakis10}. In this context 
the radio-AGN feedback (often called the ``radio'' or ``maintenance''
mode), which is thought to limit stellar mass growth in already
massive galaxies, is expected to occur only in LERAGN (\smo\ 2009).

Furthermore, it has been shown that the cosmic evolution of the
space density of various types of radio AGN is significantly different
(e.g.\ Peacock et al.\ 1985, Willott et al.\ 2001; Smolcic et al.\
2009).  Based on a study of the evolution of the radio AGN luminosity
function out to $z=1.3$, 
\smo\ et al.\ (2009) have shown that the comoving space density of 
low-luminosity radio AGN (predominantly LERAGN) only modestly declines
since $z=1.3$, while that of powerful AGN (predominantly HERAGN)
dramatically diminishes over the same cosmic time interval. 
This suggests that LERAGN and HERAGN not only represent physically
distinct galaxy populations, but also populations in different stages
of massive galaxy build-up. If this is the case, the molecular gas
masses and fractions in low- and high- excitation radio AGN are
expected to directly reflect this trend.  

We here investigate this idea by observing CO($J$=1$\to$0) emission 
of a carefully selected, representative sample of nearby ($z<0.1$) HE-
and LERAGN with CARMA.
We adopt a $\Lambda$CDM cosmology with $H_0=70$, $\Omega_M=0.3$,
$\Omega_\Lambda=0.7$.

\section{Data}
\label{sec:data}

\subsection{Sample}

We here utilize a sample of 21 Type 2 AGN at $z<0.1$ that have been
observed in X-rays (with Chandra or XMM-Newton) by \citet{evans06}.
18 out of the 21 AGN have been drawn from the 3CRR survey, adding 3
more sources (3C~403, 3C405 and Cen~A) for completeness
(see \citealt{evans06} for details). The sample properties are
summarized in \t{tab:physprops} \ (see also Tab.~1 in Evans et al.\
2006). We separate our AGN into
LERAGN (i.e.\ LINERs) and HERAGN (i.e.\ Seyferts) using standard
diagnostic tools based on optical emission line flux ratios where
possible (see \f{fig:bpt} \ and \t{tab:physprops} ;
\citealt{kauffmann03a,kewley01,kewley06}; \citealt{smo09};
\citealt{buttiglione09}).
For this we make use of the emission line fluxes extracted from high resolution spectroscopy of 3CR sources
presented in \citet[][see also Tab.~1 in \citealt{smo09}]{buttiglione09, buttiglione10, buttiglione11}. 
In cases where the relevant emission line fluxes are not available, we
make use of the galaxy type information available in the NASA Extragalactic Database\footnote{\tt
http://nedwww.ipac.caltech.edu} to separate the sources into LE- and
HE-RAGN. The sample contains 9 HERAGN and 12
LERAGN.

\begin{table*}
\begin{center}
\caption{Physical properties of the $z<0.1$ sample of radio AGN}
\label{tab:physprops}
\vskip 10pt
\begin{tabular}{|c|c|c|c|c|c|c|c|c|c|}
\hline
name & redshift & type & $L_\mathrm{178-MHz}$ & 
$L_\mathrm{0.5-10keV}/L_\mathrm{EDD}$ & stellar age & $M_*$ & $M_\mathrm{BH}$ & $M_{H_2} $\\
& && [W/Hz/srad] & & [Gyr] &  [$\mathrm{M_\odot}$] &[$\mathrm{M_\odot}$] &
 [$\mathrm{M_\odot}$]\\
\hline
3C 31 &  0.017 & Seyfert  & 9.08$\times10^{23}$ & $<2.0\times10^{-4}$   & 3&
$2.4\times10^{11}$ &  $7.8\times10^7$ &$(5.1\pm0.4)\times10^8$\\
 3C 33 &  0.060 & Seyfert & $3.95\times10^{25}$  & $1.6\times10^{-3}$ & 5 &
$1.3\times10^{11}$ & $4.8\times10^8$ &  $(3.75\pm1.5)\times10^8$\\
3C 98 &  0.030 &Seyfert & 8.75$\times10^{24}$ &$3.7\times10^{-4}$  & 2 & 
$7.9\times10^{10}$ & $1.7\times10^8$ &     $<7.8\times10^7$ \\
3C 321 &0.097 & Seyfert & 2.6$\times10^{25}$& -- & 13 &
 $7.0\times10^{11}$ &-- & $(3.3\pm0.6)\times10^9$ \\
3C 403 &  0.059 &  Seyfert & 3.5$\times10^{25}$ & $3.3\times10^{-3}$
&5&$2.4\times10^{11}$ &  $2.6\times10^8$ & $(6.6\pm1.6)\times10^8$ \\
3C 449 &  0.017 & Seyfert& 6.51$\times10^{23}$ & $<7.0\times10^{-3}$   &
3&$2.4\times10^{10}$& $5.1\times10^7$ &$(1.1\pm0.2)\times10^8$\\  
3C 452 &  0.081 &  Seyfert & 7.54$\times10^{25}$ & $3.3\times10^{-3}$
& 13 &
$4.5\times10^{11}$ &  $3.5\times10^8$ &  $8.1\times10^{8,c}$ \\ 
Cen A & 0.0008 & Sey~2$^a$ &5.4$\times10^{23}$ & $3.0\times10^{-5}$ & -- &
-- & $2.0\times10^8$ &  $1.4\times10^8$\\
3C 405 & 0.0565 & Sey~2$^a$ & 4.90$\times10^{27}$ & $8.5\times10^{-4}$ & -- &
-- & $2.5\times10^9$ & $<3.3\times10^8$ \\ 
\hline
3C 66B &  0.022 & LINER & 2.21$\times10^{24}$ &
$<4.4\times10^{-5}$  &3& $3.0\times10^{10}$ & $6.9\times10^8$ & $<7.8\times10^7$ \\
3C 84 &  0.018 &  LINER & 3.74$\times10^{24}$ & $<9.2\times10^{-6}$  &--& --
& $1.9\times10^9$ & $(2.14\pm0.02)\times10^9$  \\
3C 264 &  0.022 &  LINER  & 2.20$\times10^{24}$ &
$<1.8\times10^{-5}$  &13 & $4.4\times10^{11}$ & $7.1\times10^8$ &
$(9.3\pm1.8)\times10^7$ \\ 
3C 272.1 &  0.004 & LINER  & 3.1$\times10^{22}$  & $<8.5\times10^{-7}$
& 13 &
$2.9\times10^{11}$ & $1.5\times10^9$ &  $(9.3\pm3.2)\times10^{5,c}$ \\
3C 274 &  0.004 & LINER & 3.4$\times10^{24}$ &   $<4.3\times10^{-7}$
&2& $1.5\times10^{11}$ & $2.4\times10^9$ & $(1.65\pm0.15)\times10^7$\\ 
3C 296 &  0.025 & LINER & 1.43$\times10^{24}$ &   $<1.2\times10^{-5}$
&13& $1.1\times10^{12}$ & $1.3\times10^9$ & $<5.7\times10^7$ \\ 
3C 338 &  0.032 & LINER  & 8.63$\times10^{24}$ & $<2.0\times10^{-5}$
&13& $1.3\times10^{12}$ & $1.7\times10^9$ & $3\times10^7$\\ 
3C 388 &0.091 & LINER  & 4.29$\times10^{25}$ & -- & 9 &-- & -- & $<1.2\times10^{9,b}$

\\
3C 465 &  0.030 &  LINER  & 6.41$\times10^{24}$ & $<2.2\times10^{-4}$ &13 &
$1.0\times10^{12}$ & $2.1\times10^9$ &  $<1.95\times10^7$ \\
NGC 1265&  0.027 & LERG$^a$ & 3.39$\times10^{24}$ &
$<6.8\times10^{-6}$  &13 &$1.5\times10^{10}$ & $1.0\times10^9$ & $<5.7\times10^7$ \\
NGC 6109 & 0.0296 & LERG$^a$ &  1.86$\times10^{24}$ & -- & -- &
-- & -- & $(1.3\pm0.3)\times10^8$ \\
NGC 6251 & 0.0244 & LERG$^a$ &  1.20$\times10^{24}$ & $<2.0\times10^{-4}$ & -- &
-- & $6.0\times10^8$ & $<7.5\times10^7$\\
\hline
\end{tabular}
\end{center}
$^a$ Based on NED; LERG abbreviates ``low-excitation radio galaxy''\\
$^b$ Adopted from Saripalli et al.\ (2007), and scaled to the
cosmology used here.\\
$^c$ Not considered in our statistical analysis (see \t{tab:co} \ for details).\\
The first, second and third columns denote the source, its redshift
and AGN type. The last was 
inferred either via optical diagnostic diagrams
(see \f{fig:bpt} ), or adopted from NED. The fourth column shows the
radio continuum luminosity, adopted from Evans et al.\ (2006). The
fifth column, also adopted from
\citet{evans06}, represents the accretion
efficiency (in Eddington units) derived from X-ray observations of the
cores of the AGN (the upper limits are obtained assuming $N_H=10^{24}$~atoms~cm$^{-2}$; see
Evans et al.\ 2006 for details). The sixth column shows the stellar age
of the source based on fitting stellar population synthesis models to
the optical spectra of the sources (encompassing the $H\alpha$ portion of
the spectrum; see Tab.~5 in
Buttiglione et al.\ 2009). The seventh column shows the 
stellar mass derived using the 2MASS K-band luminosity and stellar age
(where available)
following Drory et al.\ (2004; see text for details). The second to
last column shows the black hole mass, adopted from Evans et al.\
(2006), and the last column reports the molecular gas mass obtained
from CO(1$\rightarrow$0) observations (see \t{tab:co} ) using a conversion factor
of $\alpha=M_{H_2}/L'_{CO}=1.5$~(K~\kms~pc$^2$)$^{-1}$, and assuming
$H_0=70$, $\Omega_M=0.3$,
$\Omega_\Lambda=0.7$. The horizontal lines 
separate
Seyferts (i.e.\ HERAGN; top) and LINERs (i.e.\ LERAGN; bottom). 
\end{table*}

\begin{figure*}[t!]
\includegraphics[bb=48 465 486 632,scale=1.2]{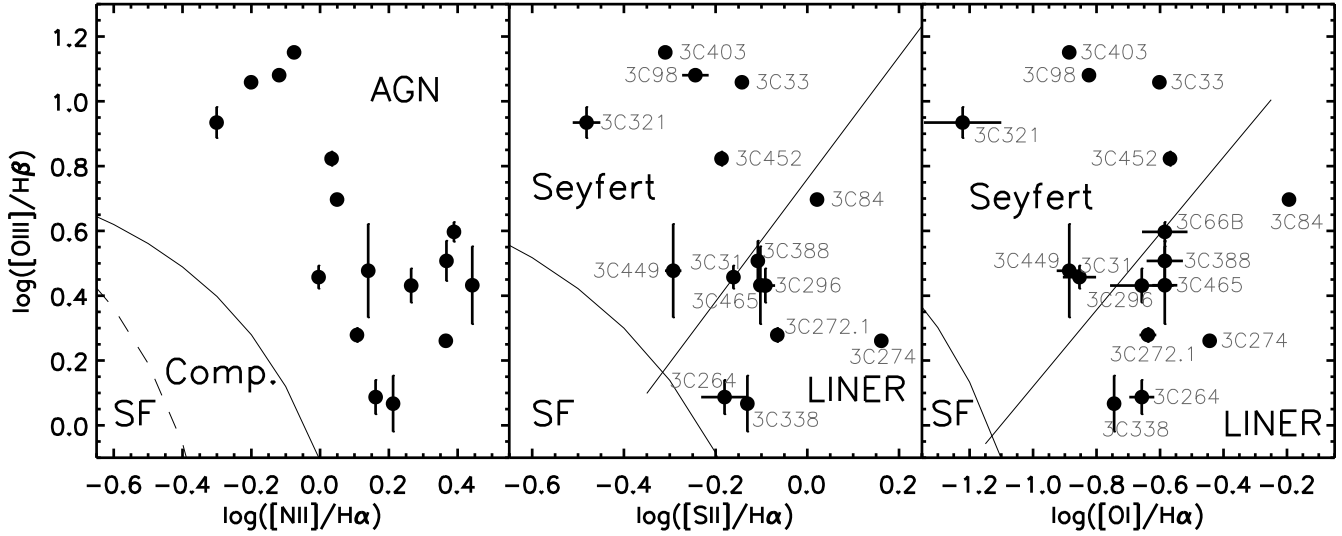}
\caption{Spectroscopic diagnostic diagrams for our 3CRR (filled dots)
  sources that separate AGN into LINERs and Seyferts. The emission
  line fluxes have been taken from Buttiglione et al.\ (2009; see
  their Tab.~1). The regions (separated following
  \citealt{kauffmann03a,kewley01,kewley06}) that are occupied by star
  forming (SF), composite (Comp.) and AGN (Seyfert and LINER) galaxies
  are indicated in the panels. Each 3CRR galaxy is also labeled.  
	\label{fig:bpt}}
\end{figure*}

\subsection{CO(1$\rightarrow$0) observations and data reduction}

At the time of observations, 8 out of the 21 Type~2 AGN in our sample
have already been detected in CO(1$\rightarrow$0). Thus, we observed
the CO(1$\rightarrow$0) transition line toward the remaining 13 AGN
using the the CARMA (Combined Array for Research in Millimeter-wave
Astronomy) Interferometer. 
Observations were performed during Summer/2009 and Spring/2010 for
about 4 to 15 hours per source (\t{tab:obs} ). All targets were
observed under good to excellent weather conditions at 3\,mm with 15
antennas (corresponding to 105 baselines)
in the two most compact, E and D configurations (2009 and 2010,
respectively). Data on two objects (3C~388 and 3C~405) had to be
discarded due to technical problems, and are excluded in the
following.  The receivers were tuned to the redshifted CO($J$=1$\to$0)
line frequencies ($\nu_{\rm rest}$=115.2712\,GHz; see \t{tab:obs} \ for exact
observing frequencies), centering them in the upper sideband.
Three bands with 15 channels of 31.25\,MHz width each were utilized.
The bands were overlapped by 2 channels to improve calibration of the
correlated dataset, leading to an effective bandwidth of 1281.25 MHz
(∼3500~\kms ) per sideband. Phase calibration was performed by
observing bright nearby radio quasars every 15\,minutes. Bandpass
calibration was performed once per track on bright quasars. Fluxes
were bootstrapped relative to planets, or monitored radio quasars if
no planet was available.
The total calibration is estimated to be accurate to 15\%. 
Data reduction was performed using the MIRIAD package. 
The CO(1$\rightarrow$0) spectra are shown in \f{fig:spec} .

\begin{table*}
\begin{center}
\caption{Summary of observations with the CARMA interferometer. }
\label{tab:obs}
\vskip 10pt
\begin{tabular}{lllcccccc}
\hline
    Source       &   RA (J2000) & DEC (J2000)  &  
    configuration & obs.\ freq. &  On-source & beam & rms/channel  \\
          &   &  & 
    & [GHz] &  time [hr] & [arcsec] & [mJy]  \\
\hline
      3C 296                       &   14 16 52.94 & +10 48 26.50 &
      D \& E   &  112.460   &  14.1 & $6.4"\times5.5"$ &
      2.1 \\
      3C 321                       &      15 31 43.45 & +24 04 19.10 &
      E & 105.079 & 4.0 & $9.8"\times7.2"$ & 3.8\\
      3C 33                         &   01 08 52.86 & +13 20 13.80 &
      D \& E  &  108.746    &  14.8 & $8.7"\times6.3"$ &
      1.8 \\
     3C 403                       &     19 52 15.80 & +02 30 24.47 &
      E & 108.849  &  5.7 & $9.0"\times6.9"$ & 4.7\\
     3C 452                       &    22 45 48.77 & +39 41 15.70 &
      D \& E &  106.634& 15.6 & $6.8"\times5.1"$ & 1.8\\
     3C 465                       &    23 38 29.52 & +27 01 55.90 &
     E & 111.914 & 8.9 & $9.3"\times6.5"$ & 3.0\\
     3C 66B                        &    02 23 11.41 & +42 59 31.38
     & E &  112.790 &4.4 & $8.7"\times6.3"$ & 5.1\\
      3C 98                         &  03 58 54.43 & +10 26
      03.00 &  E &111.914 & 4.8 & $9.3"\times6.7"$ & 3.7\\
     3C 83.1B (NGC 1265)  &   03 18 15.86 & +41 51 27.80 &   D \& E &112.241 &13.5 & $4.5"\times3.7"$ & 2.1\\
    NGC 6109                   & 16 17 40.54 & +35 00 15.10 &
    D \& E &111.957 &10.6 & $5.0"\times4.0"$ & 2.7 \\
    NGC 6251                    &  16 32 31.97 & +82 32 16.40 & E &112.526  &6.53 & $9.0"\times8.2"$ & 3.5\\
\hline
\end{tabular}
\end{center}
\end{table*}

\begin{table*}
\begin{center}
\caption{Molecular gas properties}
\label{tab:co}
\vskip 10pt
\begin{tabular}{lcccccc}
\hline
    Source       &    z         &   $S_\mathrm{cont}$ & $z_{CO}$  &
     $\Delta v_\mathrm{FWHM}$ & $I_\mathrm{CO(1\rightarrow 0)}$ & $L'_\mathrm{CO}$
    \\
          &            &   [mJy]  &  &
   [\kms ] &  [Jy \kms ] & [K \kms\ pc$^2$] \\
\hline
3C 33                         &   0.060$^*$  & $31.8\pm0.4$ &
      0.060  & $400\pm200$ &  $1.5\pm0.6$ & $(2.5\pm1.0)\times10^8$   \\
3C 66B                        &   0.022$^*$ & $113.6\pm0.8$ &
     --&--& $<2.4$ &$<5.2\times10^7$\\
3C 83.1B (NGC 1265)  &   0.027$^*$  & $40.5\pm0.4$ & --&--& $<1.2$ &$<3.8\times10^7$\\
3C 98                         &  0.030$^*$  & $8.0\pm0.4$ &
      --&--& $<1.3$ & $<5.2\times10^7$\\
3C 296                       &   0.025$^*$ & $144.4\pm0.4$ &
      --&--& $<1.4$ & $<3.8\times10^7$\\
3C 321                       &   0.097$^*$    & $9.1\pm0.6$ &
      0.097 & $320\pm70$ & $5.0\pm0.9$ & $(2.2\pm0.4)\times10^9$\\
3C 403                       &   0.059$^*$  & $4.6\pm0.5$ &
      0.059 & $350\pm100$ & $2.8\pm0.7$ & $(4.4\pm1.1)\times10^8$\\
3C 452$^+$                       &   0.081$^*$ & $44.7\pm0.6$ &--&--&--&$<5.4\times10^8$\\
3C 465                       &   0.030$^*$  & $20.2\pm0.1$ &
      --&--& $<0.3$ & $<1.3\times10^7$\\
NGC 6109                    &   0.0296$^{**}$        & $17.7\pm0.3$ &
    0.0301 &  $230\pm50$ & $2.2\pm0.4$ & $(8.8\pm1.7)\times10^7$\\
NGC 6251                    &   0.0244$^{**}$        & $624.9\pm0.6$&
    --&--& $<1.9$ &$<5.0\times10^7$\\
\hline
Cen A$^e$ & 0.0008$^{**}$ & -- & -- & -- & -- & $9.4\times10^7$\\
3C 272.1 (M84) $^{b,c}$ & 0.004$^*$ & -- & 0.0028 & 200 & $1.8\pm0.6$ & $(6.2\pm2.1)\times10^5$\\
3C 274 (M87)$^b$ & 0.004$^*$ & -- & 0.0035 & 200 & $20\pm2$ &  $(1.1\pm0.1)\times10^7$\\
3C 31$^a$ & 0.017$^*$ & -- & 0.0169& 450 & $27\pm2$ &
$(3.4\pm0.2)\times10^8$ \\
3C 449$^b$  & 0.017$^*$ & -- & 0.0169 & 500 & $6\pm1$ & $(7.6\pm1.3)\times10^7$\\
3C 84 (NGC 1275)$^a$ & 0.018$^*$ &  -- & 0.0176 & 
200 & $104\pm1$ & $(1.43\pm0.01)\times10^9$\\
3C 264$^b$ & 0.022$^*$ & -- & 0.02 & 200 & $3.5\pm0.7$ & $(6.2\pm1.2)\times10^7$\\
3C 338$^d$ & 0.032$^*$ & -- & 0.030 & -- & 0.495 &  $2.0\times10^7$\\
3C 405$^a$ & 0.0565$^{**}$ & -- & -- & -- & $<1.5$ & $<2.2\times10^8$ \\
3C 388                       &   0.091         & --  &--&--& -- & -- \\
\hline
\end{tabular}
\end{center}
The columns show the source, its redshift, the observed continuum
flux density ($S_\mathrm{cont}$), the redshift based on the
CO(1$\rightarrow$0) emission line ($z_\mathrm{CO}$), the line width at half-maximum
($v_\mathrm{FWHM}$), the CO line intensity ($I_\mathrm{CO(1\rightarrow
  0)}$), and luminosity ($L'_\mathrm{CO}$; see
eq.~4 in Evans et al.\ 2005). For sources in which the CO
line was not detected we report 
$3\sigma$ upper limits, computed assuming $\Delta
v_\mathrm{FWHM}=300$~\kms .\\
$^*$~adopted from Buttiglione et al.\ (2009) \\
 $^{**}$~adopted from Evans et al.\ (2006) \\
$^+$~Due to the strong contribution of complex, steeply sloped mm continuum 
emission from extended radio jets to the mm
emission of this source, the continuum was fitted over only 33
channels, where the jet contribution is estimated to be
small after deconvolution. Due to this uncertainty, however, we 
do not consider this source in our statistical analysis. \\
$^a$~$z_\mathrm{CO}$, $\Delta v_\mathrm{FWHM}$ and
 $I_\mathrm{CO(1\rightarrow 0)}$ are adopted from Evans et al.\
 2005. $L'_\mathrm{CO}$ was computed
 using the cosmology adopted here. \\
 $^b$~$z_\mathrm{CO}$, $\Delta v_\mathrm{FWHM}$ and $I_\mathrm{CO(1\rightarrow 0)}$ are
 taken from Flaquer et al.\ (2010). Given that their observations were
 conducted with  the IRAM 30~m telescope, we take $1~\mathrm{K}  =
 4.95$~Jy, and compute $L'_\mathrm{CO}$ 
 using the cosmology adopted here. \\
$^c$ tentative detection\\
$^d$~$I_\mathrm{CO(1\rightarrow 0)}$ adopted from Leon et al.\
(2001). \\
$^e$~adopted from Eckart et al.\ (1990), and scaled to the cosmology
used here.\\
\end{table*}

\begin{figure*}
\includegraphics[angle=-90, scale=0.6]{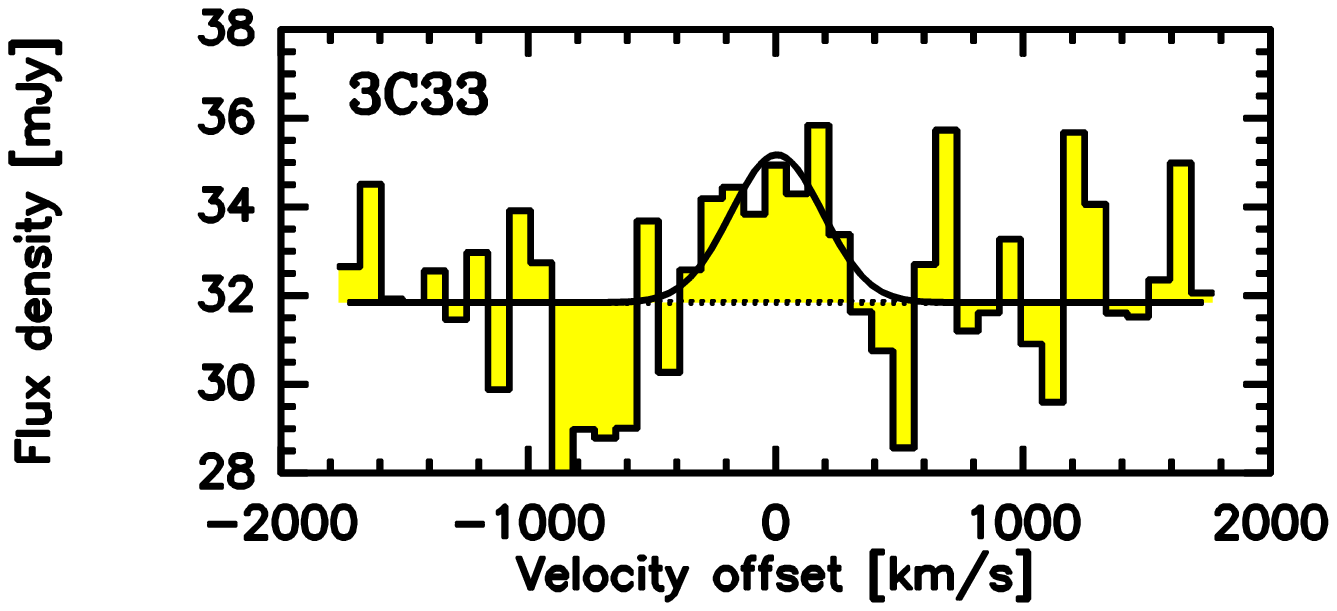}
\includegraphics[angle=-90, scale=0.6]{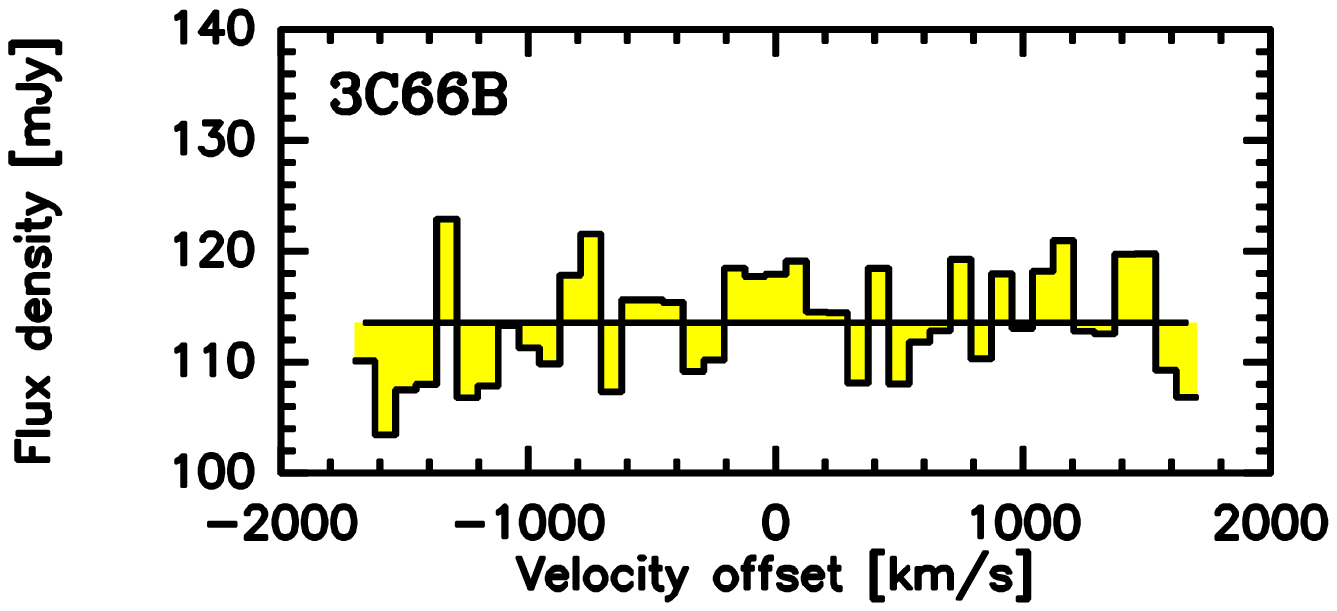}\\
\includegraphics[angle=-90,scale=0.6]{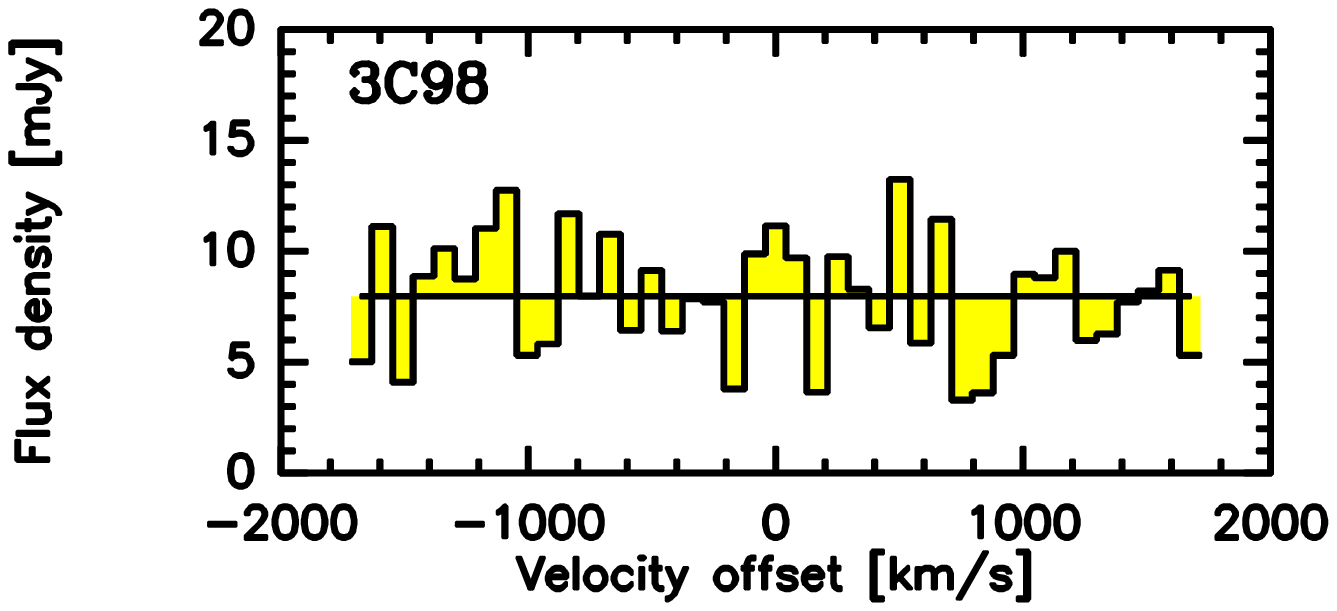}
\includegraphics[angle=-90, scale=0.6]{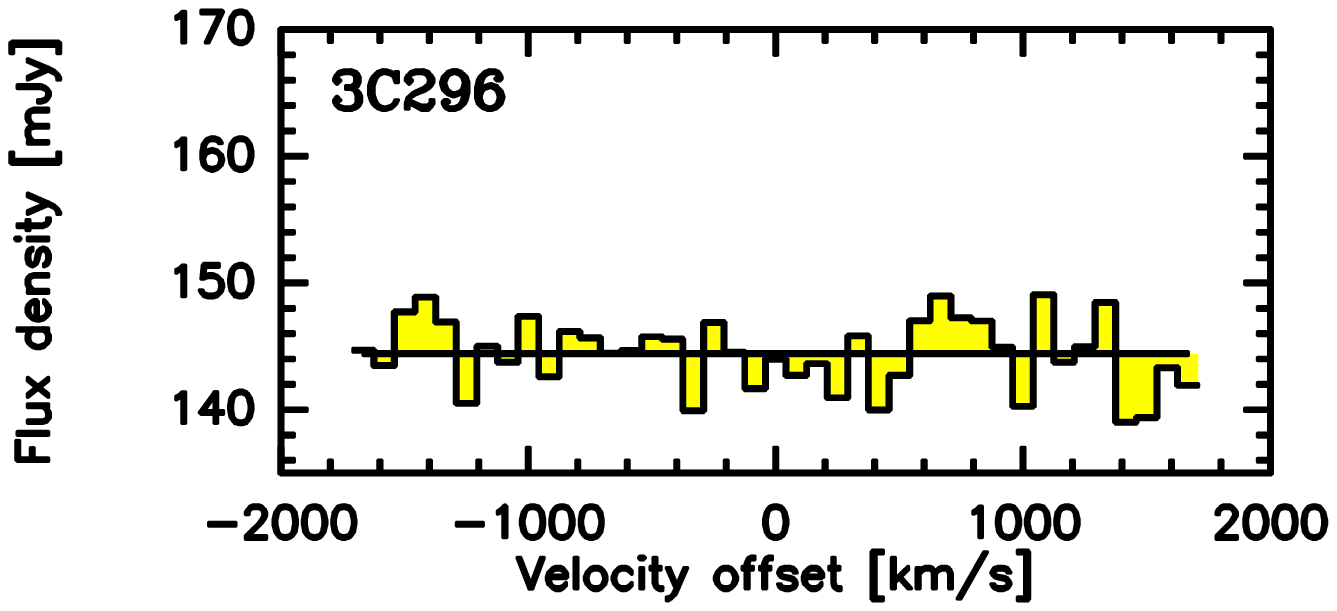}\\
\includegraphics[angle=-90, scale=0.6]{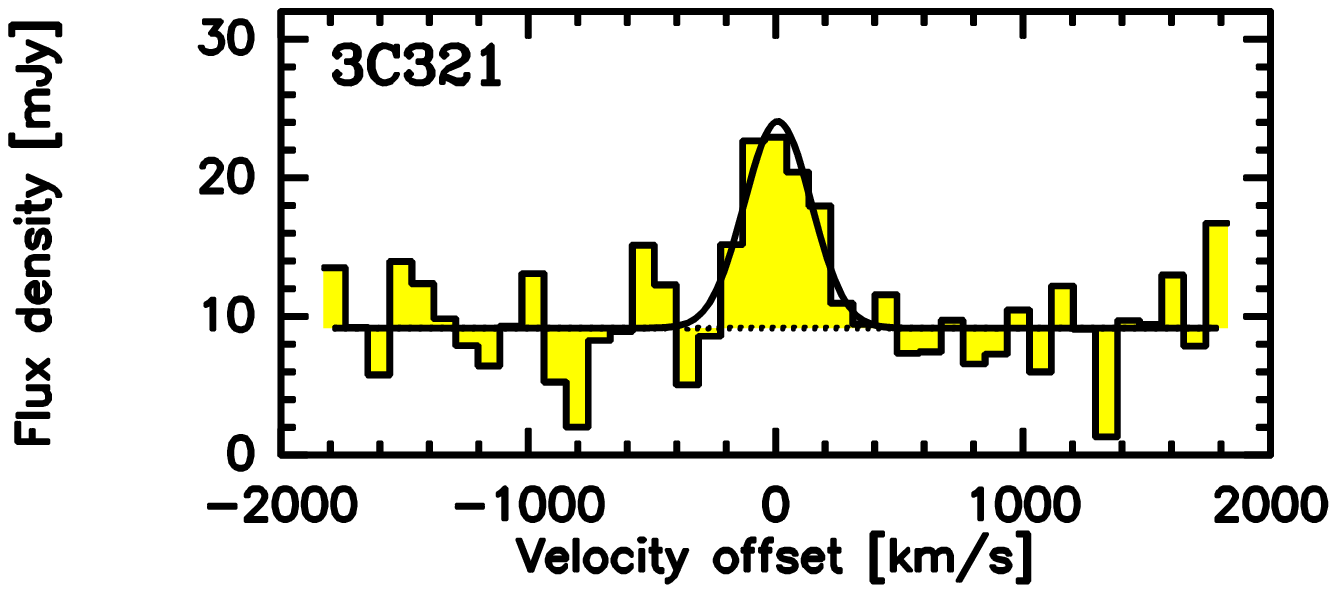}
\includegraphics[angle=-90, scale=0.6]{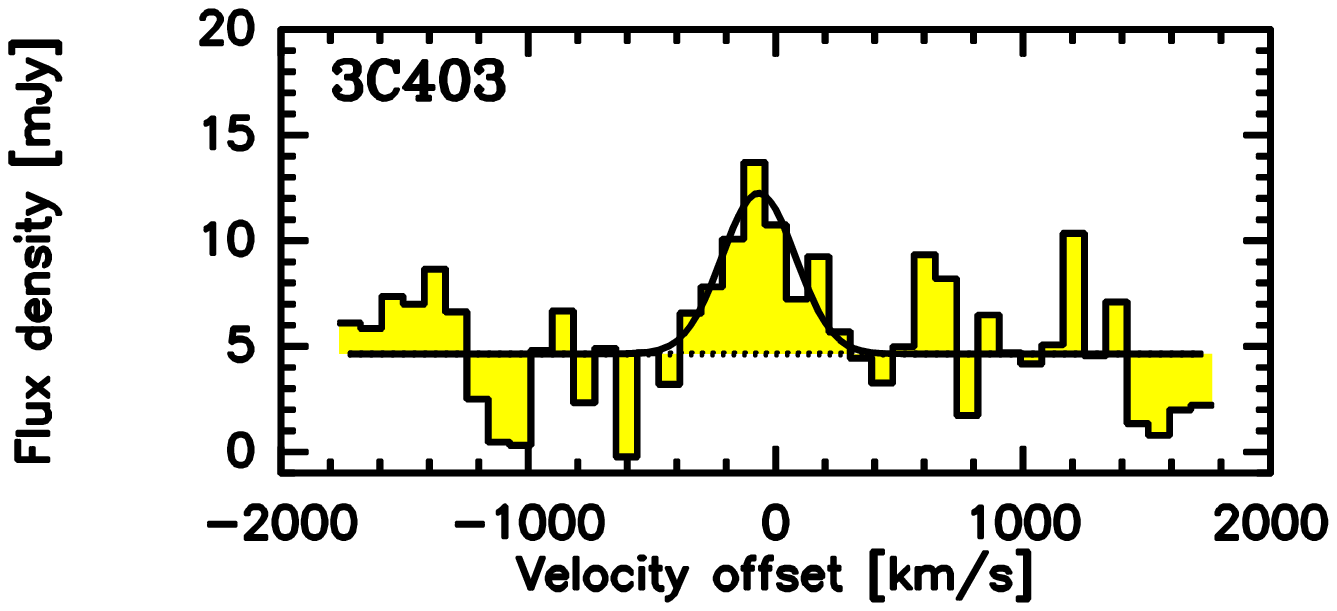}\\
\includegraphics[angle=-90, scale=0.6]{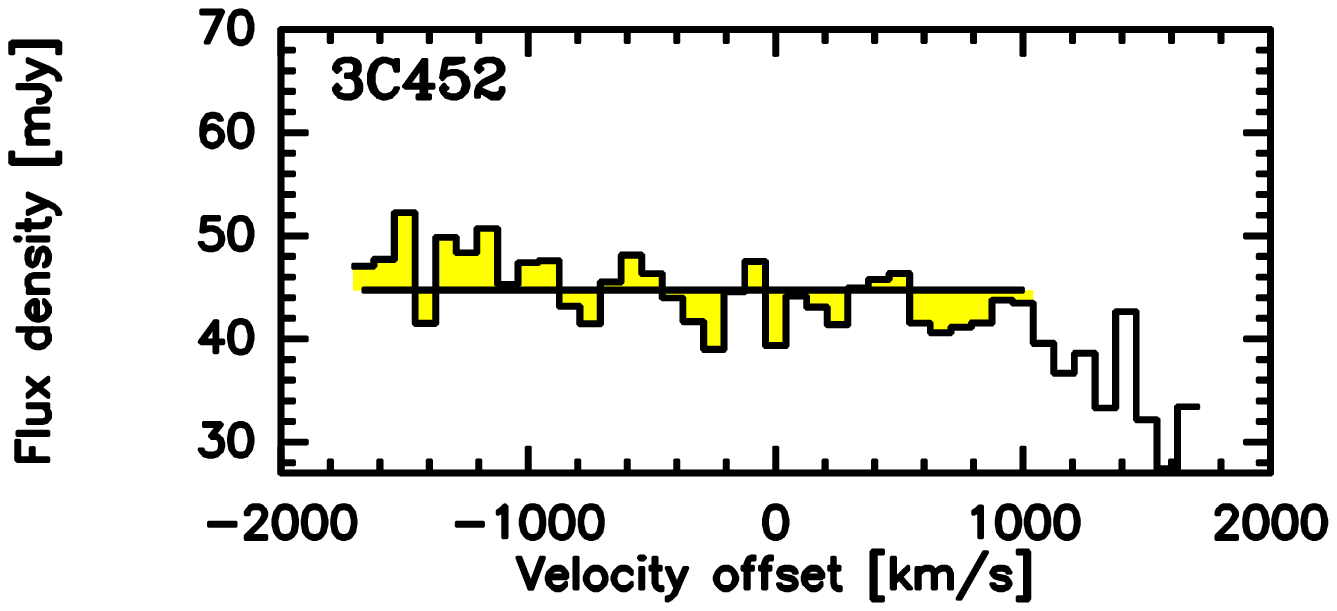}
\includegraphics[angle=-90, scale=0.6]{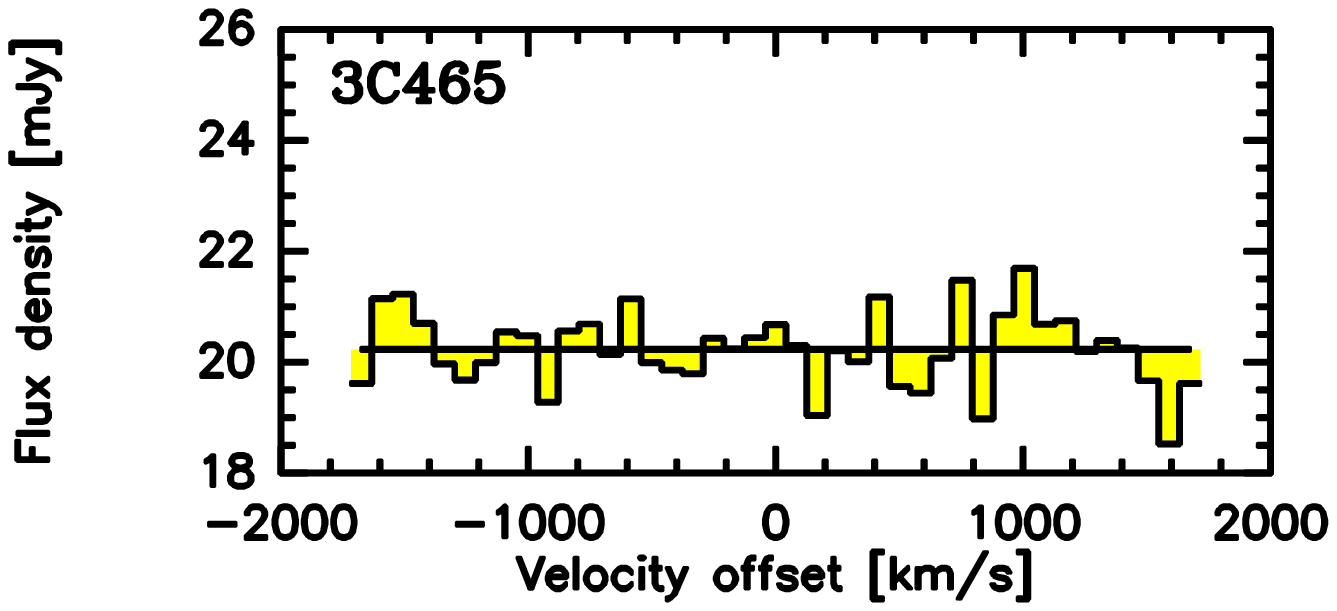}\\
\includegraphics[angle=-90, scale=0.6]{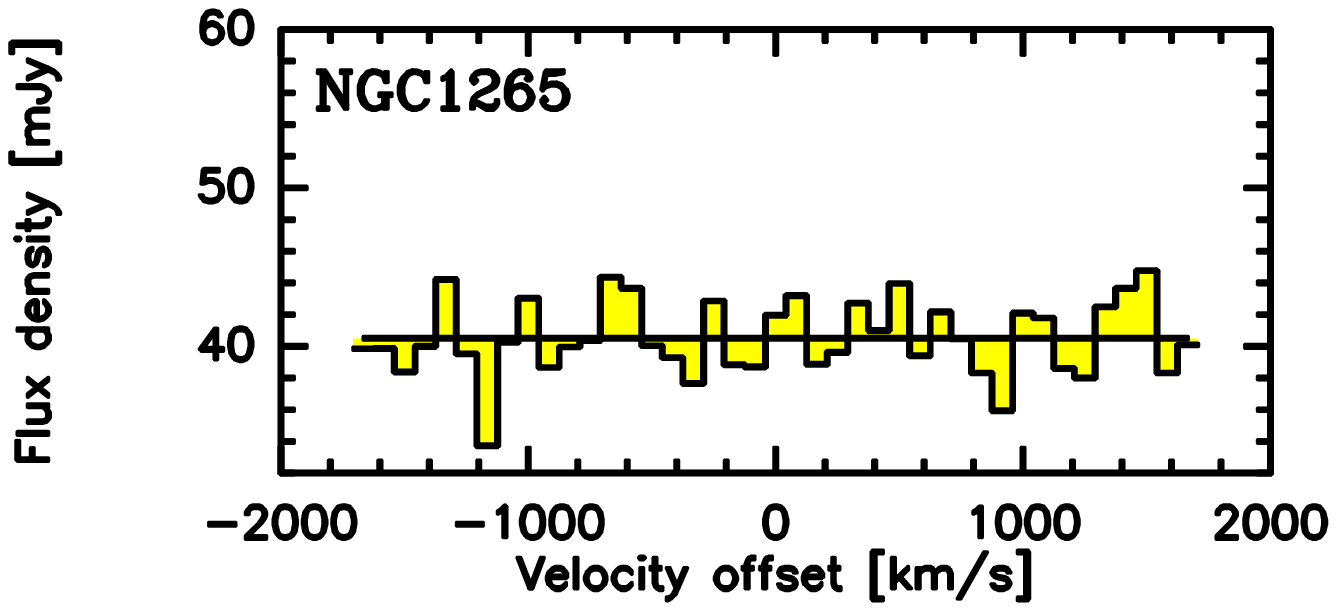}
\includegraphics[angle=-90, scale=0.6]{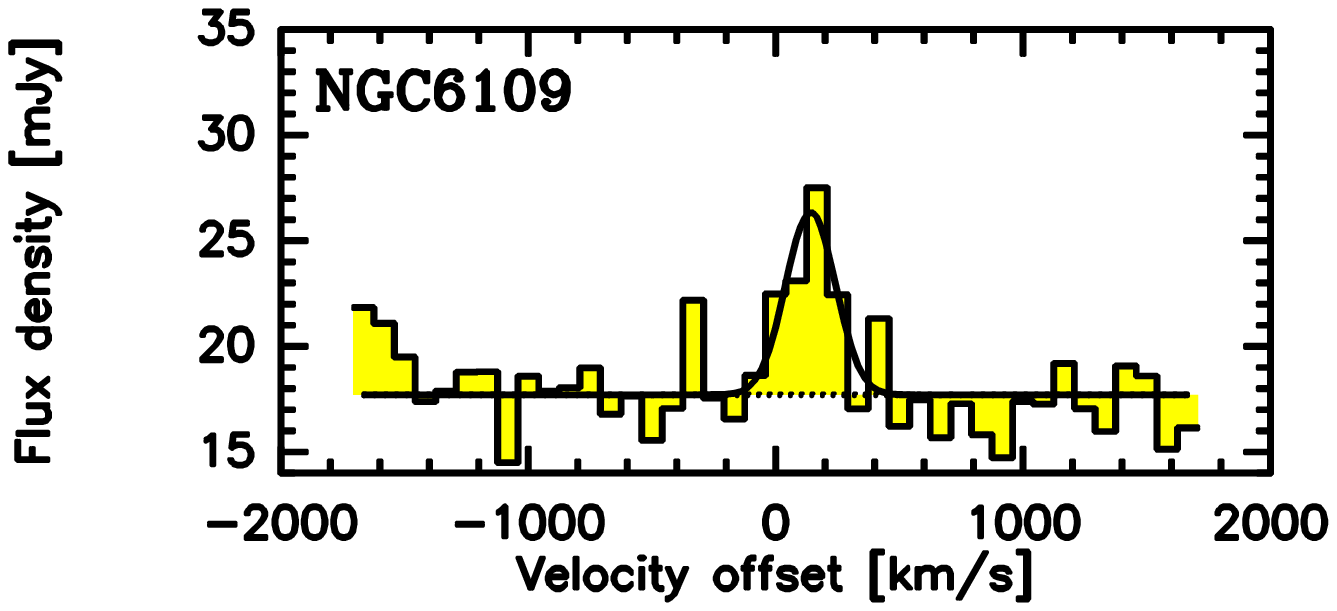}\\
\includegraphics[angle=-90, scale=0.6]{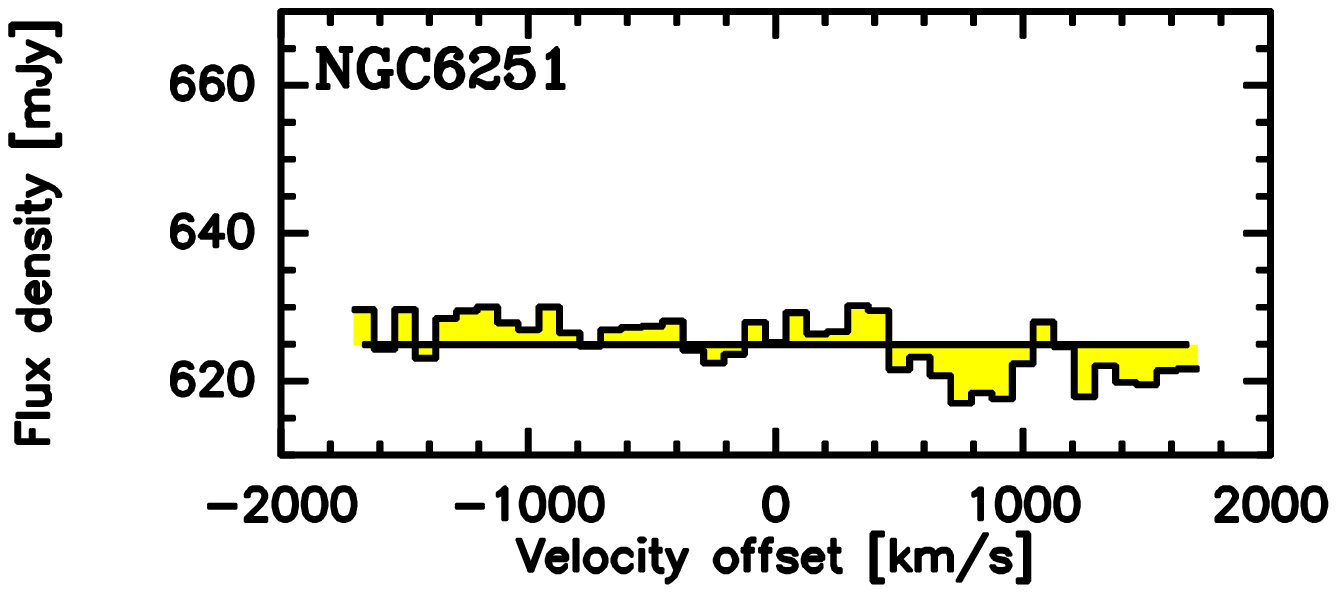}
\caption{CO(1$\rightarrow$0) spectra (histograms) of the 11 radio galaxies in our sample
  observed with CARMA. The data are shown at a resolution of
  31.25\,MHz (83--89\,\kms ). The black curves are Gaussian fits to
  the line profiles and rest-frame 2.6\,mm continuum emission. } \label{fig:spec}
\end{figure*}

\section{Results}
\label{sec:results}

\subsection{CO Data}

CO(1$\rightarrow$0) has been detected in 4 (3C33, 3C321, 3C403, and
NGC6109) out of the 11 galaxies in our CARMA-CO sample (see
\f{fig:spec} ).  To parameterize the emission lines detected in these
four galaxies we fit Gaussian profiles to the line and underlying
continuum emission (see \t{tab:co} \ for line/continuum properties).

Two of our four CO detected sources (3C321 and 3C403) have recently
been detected in the CO(1$\rightarrow$0) transition by Flaquer et al.\
(2010) using the IRAM 30~m telescope. The line parameters reported by
Flaquer et al.\ are in good agreement with ours.

$3\sigma$ upper limits for CO(1$\rightarrow$0) non-detections are
determined by assuming a line width of 300~\kms, corresponding to the
average width of the detected lines.
We further complement \t{tab:co} \ with data from literature for the 8
sources with already existing CO(1$\rightarrow$0) detections, and the
2 sources (3C~388 and 3C405) that had to be excluded from our sample.

\subsection{Ancillary Data}

We summarize the physical properties of the 21 sources in
our low- and high-excitation radio AGN sample in \t{tab:physprops} .
We adopt the 178~MHz luminosities, accretion efficiencies, and black
hole masses from Evans et al.\ (2006). The stellar ages of our
sources, taken from Buttiglione et al.\ (2009), were derived by
fitting stellar population synthesis models to the ($H\alpha$ portion
of the sources') optical spectra. Combining the stellar ages with
2MASS K-band luminosities (where available) we computed the stellar
masses of our sources following Drory et al.\ (2004). Drory et al.\
have parameterized the mass-to-light ratio in K-band as a function of
stellar age (see their Fig.~1) using simple stellar population models
(Maraston 1998), and a Salpeter initial mass function. The total
systematic uncertainty of such a derived mass-to-light ratio is
estimated to be $\sim25-30$\%. Lastly, from the CO(1$\rightarrow$0)
luminosities inferred for our sources (see \t{tab:co} ) we estimated
the molecular ($H_2$) mass using a conversion factor of
$\alpha=M_\mathrm{H_2}/L'_\mathrm{CO}=1.5$~(K~\kms~pc$^2$)$^{-1}$
\citep{evans05}.

We find systematic differences in the average black hole and host
galaxy properties of the low- and high- excitation sources (i.e.\
LINERs and Seyferts, respectively) in our sample (\t{tab:physprops}
). This is illustrated in \f{fig:physprops} , where we also indicate
the average properties of our low- and high-excitation radio AGN,
computed using the ASURV statistical package and assuming log-normal
distributions in luminosity and mass. The average properties are
specifically given in \t{tab:averageprops} .\footnote{It should be
  kept in mind that in ASURV there is an implicit assumption that the
  censored data follow a similar distribution to that of the measured
  population. If this is not the case, "average" values calculated by
  ASURV will be generally biased upwards (as our upper limits
  typically lie towards the bottom-end of the distribution). Note
  however that, if this were the case, it would not change, but only
  strengthen the results presented here.}
\begin{table*}
\begin{center}
\caption{Average properties of the $z<0.1$ high- and low- excitation radio AGN}
\label{tab:averageprops}
\vskip 10pt
\begin{tabular}{|c|c|c|c|c|c|}
\hline
AGN & $L_\mathrm{178-MHz}$ & stellar age & $M_*$ & $M_\mathrm{BH}$ & $M_{H_2} $\\
type &[W/Hz/srad] & [Gyr] &  [$\mathrm{M_\odot}$] &[$\mathrm{M_\odot}$] &
 [$\mathrm{M_\odot}$]\\
\hline
HERAGN & $(7.2\pm4.9)\times10^{24}$ & $6.3\pm1.6$ & $(1.7\pm0.7)\times10^{11}$ &
$(2.5\pm1.0)\times10^{8}$ & $(2.9\pm1.2)\times10^8$\\
\hline
LERAGN & $(2.5\pm1.1)\times10^{24}$ & $10.2\pm1.4$ & $(2.4\pm1.4)\times10^{11}$ &
$(1.3\pm0.2)\times10^{9}$ & $(4.3\pm1.9)\times10^{7,*}$\\
\hline
\end{tabular}
\end{center}
$^*$~The given limit was computed excluding the tentative CO detection in
3C~272.1 (see \t{tab:co} ). Including the gas mass for this
source yields an average of $(1.8\pm1.5)\times10^7$~\msol .
\end{table*}
\begin{figure*}
\includegraphics[bb= 54 460 486 662,scale=1.15]{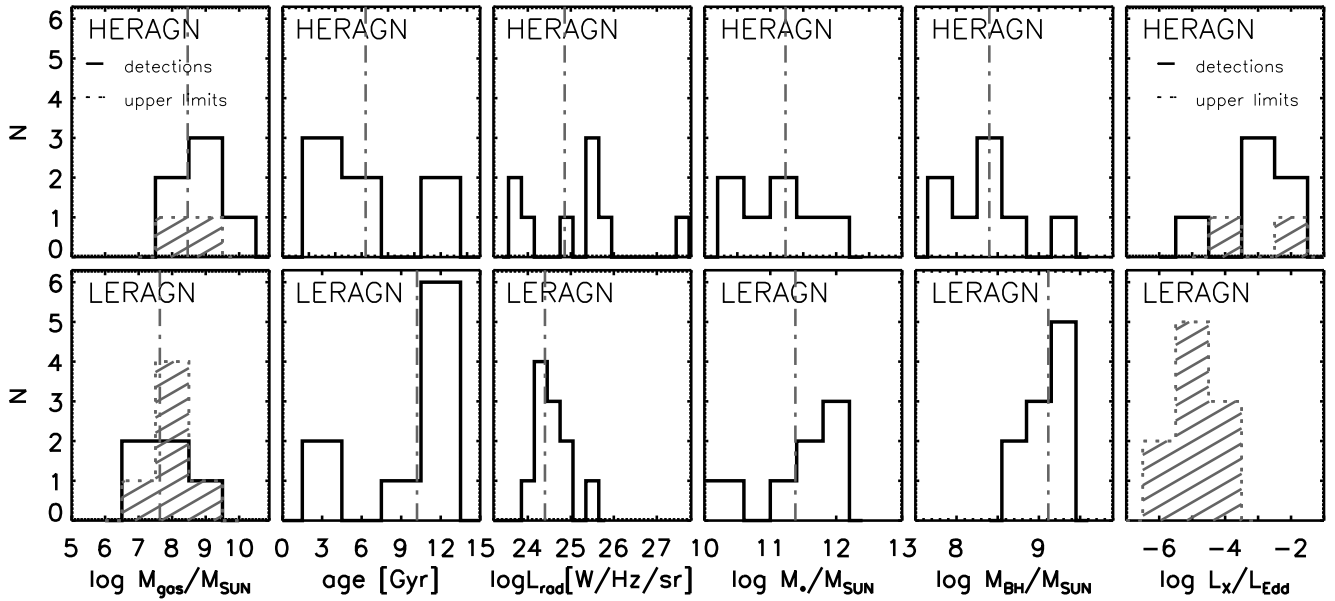}
\caption{ Distribution of physical properties of high- and low-excitation radio AGN (HEARGN
  and LERAGN, resp.), drawn from \t{tab:physprops} \ (excluding the 
  tentative CO detection in 3C~272.1). Average values
  (given in \t{tab:averageprops} ) 
  are indicated by vertical dot-dashed lines}
	\label{fig:physprops}
\end{figure*}
Compared to LERAGN, HERAGN on average have a factor of $\sim3$ higher
radio continuum luminosities, significantly higher accretion
efficiencies, but about an order of magnitude lower mass central black
holes. Furthermore, their host galaxies have about a factor of 1.5
younger stellar populations and stellar masses, but about a factor
of $\sim7$ higher molecular gas masses. As discussed in the next
section, this is consistent with the idea that high- and low-
excitation radio AGN form two physically distinct populations of
galaxies that reflect different phases of massive galaxy formation.

\section{Discussion and Summary}
\label{sec:discussion}

Our main result is that HERAGN have systematically higher molecular
gas masses (a factor of $\sim7$; see \t{tab:averageprops} ),
compared to LERAGN. 
Flaquer et al.\ (2010) have found a similar trend by
dividing their sample ($\sim50$ radio AGN observed with the IRAM 30m telescope,
partially overlapping with our sample) into FR class I and II
objects. They find that the molecular gas mass in FR~IIs is a factor
of $\sim4$ higher than that in FR~Is. The FR class can be taken to
roughly correspond to the low- and high- excitation
classification.\footnote{Almost all FR I –- low power -– radio
galaxies are LERAGN, while optical hosts of FR IIs, which are
typically more powerful than FR Is (Fanaroff \& Riley 1974; Owen 1993;
Ledlow \& Owen 1996), usually have strong emission lines. Note however
that the correspondence between the FR class and the presence of
emission lines is not one-to-one.}  Flaquer et al.\ (2010) have,
however, concluded that the systematic differences they find are
likely a result of a Malmquist bias, i.e.\ simply due to a
systematically higher redshift of their FR-II sources. Although our
HERAGN lie on average at a slightly higher redshift, compared to our
LERAGN (0.046 vs.\ 0.030, resp.) in the following we argue that the
systematic differences we find in molecular gas mass are not due to a
Malmquist bias.

Mori\'{c} et al.\ (2010) have shown that the redshift distributions of
carefully selected samples of radio-selected LINERs and Seyferts are
approximately the same (see their Fig.~6). This eliminates Malmquist
bias from their results.  They find that the detection fraction in the
FIR is significantly lower for LINERs than for Seyferts (6.5\% vs.\
22\%, resp.) in their sample. Assuming that the star formation law
parameterized by $L'_{\rm CO}$ (as a proxy for total gas mass) and
$L_{\rm FIR}$ (as a proxy for star formation rate; e.g., Kennicutt
1998; Solomon \& Vanden Bout 2005; Bigiel et al.\ 2008), on average,
correctly represents the star formation properties of these samples
(as confirmed by the CO/FIR luminosities of the IRAS detected
sources analyzed here; see \f{fig:cofir}  ), the lower 
average FIR luminosity
in low-excitation sources (i.e.\ LINER) implies lower gas masses than in
high-excitation (i.e.\ Seyfert) types of galaxies. 
A similar result is obtained based on
average (optically derived) star formation rates\footnote{Mori\'{c} et
al.\ (2010) derived star formation rates for each galaxy in their
sample via stellar population synthesis model fitting to the SDSS
photometry of the host galaxy (see also \smo\ et al.\ 2008).}, 
suggesting that those in LINERs are by about a factor of 3 lower than
in Seyferts in a redshift-matched sample. 
These findings suggest that the systematic differences in molecular
gas mass in high- and low-excitation radio AGN are physical, and not
due to Malmquist bias.

\begin{figure}
\includegraphics[bb=94 400 410 712,scale=1.5,width=\columnwidth]{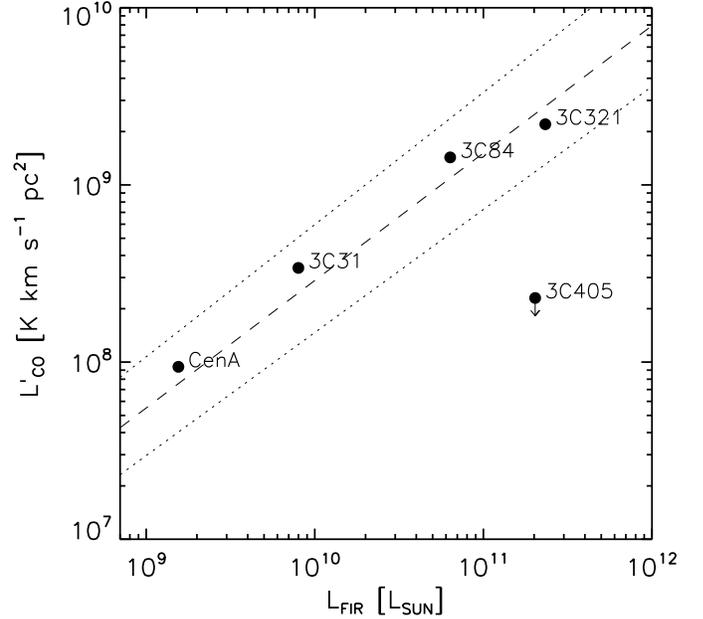}
\caption{ CO vs.\ FIR luminosity for our local AGN sources detected
  with IRAS. The lines represent the $L'_\mathrm{CO} - L_\mathrm{FIR}$ correlation derived by
  \citet{riechers06}. }
	\label{fig:cofir}
\end{figure}

The systematically higher molecular gas masses that we find in HERAGN,
relative to LERAGN in our $z<0.1$ radio AGN sample, are in excellent
agreement with the systematic differences in various properties of
high- and low- excitation radio AGN, both on pc- and kpc- galaxy
scales (see \s{sec:intro} \ and \t{tab:physprops} ).

We find that, on average, HERAGN have lower stellar masses and stellar
ages compared to LERAGN (see \t{tab:averageprops} ; see also \smo\
2009). This is consistent with HERAGN and LERAGN being green
valley and red sequence sources, respectively. Furthermore, we show
that HERAGN have on average higher radio luminosities than LERAGN,
consistent with the results presented in Kauffmann et al.\
(2008). 
Kauffmann et al.\ have shown that the fraction of radio AGN with
strong emission lines in their spectra significantly rises beyond
$\sim10^{25}$~\wh . In general, the comparison of the black hole and
host galaxy properties inferred for our 21 $z<0.1$ AGN with much
larger 
samples of radio AGN (Kauffmann et al.\ 2008; \smo\ 2009) suggests
that our AGN sample is representative of high- and low-excitation
radio AGN in the nearby universe.

From the average stellar masses that we infer for our high- and low
excitation sources we extrapolate that they occupy
$\sim3\times10^{13}$~\msol\ and $\sim5\times10^{14}$~\msol\ halos,
respectively \citep[e.g.][]{behroozi10, moster10}. Compared to the
systematic molecular gas mass difference, this yields an even more
dramatic discrepancy of more than 2 orders of magnitude in the average
molecular gas fractions in HE- ($\sim10^{-5}$) and LE-RAGN
($\sim9\times10^{-8}$). The discrepancy remains significant (about an
order of magnitude) if the average gas-to-stellar mass fraction (which
can be interpreted as star formation efficiency) is considered.

On small scales, the average black hole accretion efficiencies in HE-
and LE-RAGN suggest different supermassive black-hole accretion
mechanisms (standard disk accretion of cold gas in HERAGN vs.\ Bondi
accretion of hot gas in LERAGN; see Evans et al.\ 2006; Hardcastle et
al.\ 2006). Furthermore, the higher black hole masses in LERAGN
suggest a later evolution stage of their host galaxies, compared to
that of HERAGN. This is further strengthened by the higher stellar
masses in LERAGN, as well as older stellar ages, and less massive gas
reservoirs. In the blue-to-red galaxy formation picture, blue gas rich
galaxies are thought to transform into read-and-dead gas-poor
galaxies, the stellar populations in the host galaxies of HERAGN are
expected to be younger and have lower masses, while their molecular
gas reservoirs -- fueling further stellar mass growth -- are expected
to be higher than those in LERAGN. This is in very good agreement with
the results presented here. Thus, in summary, our results strengthen
the idea that low- and high-excitation radio AGN form two physically
distinct galaxy populations that reflect different stages of massive
galaxy formation.

\acknowledgments The authors thank F.~Bertoldi  and K.~Knudsen for insightful
discussions. The research leading to these results has received funding from the European Union's Seventh Framework programme under grant agreement 229517. DR acknowledges support from NASA through an
award issued by JPL/Caltech, and from NASA
through Hubble Fellowship grant HST-HF-51235.01 awarded by the Space
Telescope Science Institute, which is operated by the Association of
Universities for Research in Astronomy, Inc., for NASA, under contract
NAS 5-26555.  Support for CARMA construction was derived from the
Gordon and Betty Moore Foundation, the Kenneth T. and Eileen L. Norris
Foundation, the James S. McDonnell Foundation, the Associates of the
California Institute of Technology, the University of Chicago, the
states of California, Illinois, and Maryland, and the National Science
Foundation. Ongoing CARMA development and operations are supported by
the National Science Foundation under a cooperative agreement, and by
the CARMA partner universities.

{}

\end{document}